 \documentclass[aps,prl,notitlepage,superscriptaddress,twocolumn ]{revtex4}
\usepackage{graphicx,subfigure,epsfig}
\usepackage{dcolumn}
\usepackage{amssymb}
\usepackage{times}
\usepackage{amsmath}
\usepackage{amsfonts}
\usepackage{mathrsfs}
\usepackage{setspace}
\usepackage{latexsym}
\usepackage{bbm}
\usepackage{float}
\usepackage{flafter}
\usepackage{bm}
\usepackage{epstopdf}
\usepackage{hyperref}
\usepackage{color}
\usepackage{multirow}

\begin{document}

\title{Spin-Orbital Density Wave and a Mott Insulator in a Two-Orbital Hubbard Model on a Honeycomb Lattice}

  \author{Zheng Zhu}
\affiliation{Department of Physics, Massachusetts Institute of Technology, Cambridge, MA, 02139, USA}
\affiliation{Department of Physics and Astronomy, California State University, Northridge, CA, 91330, USA}
\affiliation{Department of Physics, Harvard University, Cambridge, MA, 02138, USA}
 \author{D. N. Sheng}
 \email[\href{mailto:donna.sheng1@csun.edu}{donna.sheng1@csun.edu}]{}
 \affiliation{Department of Physics and Astronomy, California State University, Northridge, CA, 91330, USA}
 \author{Liang Fu}
 \email[\href{mailto:liangfu@mit.edu}{liangfu@mit.edu}]{}
 \affiliation{Department of Physics, Massachusetts Institute of Technology, Cambridge, MA, 02139, USA}
\date{\today}

\begin{abstract}
Inspired by recent discovery of correlated insulating states in twisted bilayer graphene (TBG), we study a two-orbital Hubbard model on the honeycomb lattice with two electrons per unit cell. Based on the real-space density matrix renormalization group (DMRG) simulation, we identify a metal-insulator transition around $U_c/t=2.5\sim3$. In the vicinity of $U_c$, we find strong spin/orbital density wave fluctuations at commensurate wavevectors, accompanied by weaker incommensurate charge density wave (CDW) fluctuations. The spin/orbital density wave fluctuations are enhanced with increasing system sizes, suggesting the possible emergence of long-range order in the two dimensional limit.  At larger $U$,  our calculations indicate a possible nonmagnetic Mott insulator phase without spin or orbital polarization. Our findings offer new insights into correlated electron phenomena in twisted bilayer graphene and other multi-orbital honeycomb materials.
\end{abstract}

\maketitle
\emph{Introduction.}---The metal-insulator transition (MIT) driven by electron repulsion is one of the long-standing issues in condensed matter physics \cite{Mott,Mott1990,Imada1998}. In particular, the systems with multiple orbits may harbor different exotic phases and novel phase transitions due to the interplay between orbit, spin and charge degrees of freedom.

Recently, correlated insulators were discovered at low temperature in twisted bilayer graphene (TBG) with an integer number of electrons per superlattice unit cell ~\cite{Cao1,Cao2,Yankowitz2018}. These insulating states are found near the ``magic'' twist angle ~\cite{Cao1,Cao2} or under pressure ~\cite{Yankowitz2018}, where the lowest mini-band of TBG has a very narrow bandwidth. As a result of the reduced kinetic energy, the correlation effect becomes significant and most likely is the driving force for the MIT. The nature of the insulating states is now under intensive study ~\cite{Fu2018a, Yang,Senthil,Kivelson,Vafek2, Pizarro,Zhang,PALee,Xu,Irkhin,Rademaker,Kuroki,Fernandes,Fu2018c,You}. A paradigmatic theoretical model for MIT is the Hubbard model and its various generalizations to include multiple orbits and extended interactions.
For TBG, a two-orbital extended Hubbard model on the honeycomb lattice ~\cite{Fu2018a, Fu2018b} was recently derived from symmetry-adapted maximally-localized Wannier states of the low-energy bands ~\cite{Fu2018b,Vafek}.
Since these low-energy bands are well separated from higher bands by a sufficiently large energy gap ~\cite{Cao2016,Koshino2017}, this two-orbital extended Hubbard model is a reasonable and adequate starting point for studying correlated electron phenomena in TBG~\cite{Fu2018b, YuanFu}. In this model, the two sublattices of the emergent honeycomb lattice correspond to Wannier states centered at AB and BA stacking regions, and the two orbits are associated with states from opposite valleys of graphene. 

In a completely different context, the two-orbital $SU(4)$-symmetric Hubbard model  on the honeycomb lattice was recently proposed for $d^1$ transition metal compounds with edge-sharing anion octahedra such as $\alpha$-ZrCl$_3$ \cite{Yamada}, where spin and orbital combined together describe the $j=3/2$ quartets that emerge in the strong spin-orbit-coupling limit.

In addition to its potential relevance to real materials, it is also of fundamental importance to study the physics beyond $SU(2)$. In contrast to large-$S$ representations of $SU(2)$, the quantum fluctuations in $SU(N)$ increase with $N$ and thus opening new realm to explore more intriguing physics. In this work, we study the two-orbital Hubbard model with $SU(4)$ symmetry on the honeycomb lattice at quarter filling, i.e., with two fermions per unit cell.
 Based on large-scale density matrix renormalization group~\cite{White1992} (DMRG) simulations, we examine the nature of the ground state  with the ratio of on-site Coulomb repulsion $U$ and bandwidth $t$. We identify a MIT occurs around $U_c/t=2.5\sim3$  and a nonmagnetic Mott insulator in the large $U$ regime.
Interestingly, in the vicinity of $U_c$, we find a small region where both spin/orbital density wave (SODW) and charge density wave (CDW) fluctuations become strong.  The SODW fluctuations have commensurate wavevectors near $U_c$, and become enhanced when increasing system size, indicating long-range SODW order in the two dimensional (2D) limit. In contrast, the CDW fluctuations  have incommensurate wavevectors and are unstable against increasing system size. The SODW is robust against perturbations including  nearest neighbor interactions or lightly doping. We do not find any sign of spin/orbital polarization in the ground state.

Our findings of the nonmagnetic Mott insulator at large $U$ is consistent with previous studies of the $SU(4)$ Heisenberg model on honeycomb lattice \cite{Mila,Penc}, which is the effective Hamiltonian of our Hubbard model in $U/t \rightarrow \infty$ limit. Meanwhile, our finding of the SODW near MIT at intermediate $U$ is a new result. Our work shows that the phase diagram of $SU(4)$ Hubbard model is distinctly different from SU(2) case \cite{Meng2010,Chen2014,Sorella2012,Assaad2013,Hassan2013,Toldin2015,Otsuka2015, Shirakawa2015,Szasz2018}, where a magnetic ordered phase often appears in large $U$ limit,  a direct phase transition\cite{Sorella2012,Assaad2013,Hassan2013,Toldin2015,Otsuka2015} or
an intermediate  phase\cite{Meng2010,Chen2014,Shirakawa2015,Szasz2018}  are found  close to the MIT. It is also distinguished from half-filled  $SU(2N\geq4)$ case, where a direct  transition from semimetal to valence bond solid was identified\cite{Zhou2016,Li2019}.

\emph{Model and Method.}---We consider a  $SU(4)$ symmetric two-orbital Hubbard model on the honeycomb lattice. The model is given by
 \begin{align} \label{Model}
H &= H_0 + H_\mathrm {int},\\
H_0 &=  - t\sum\limits_{\langle {i,j} \rangle \sigma } {\sum\limits_{\alpha  = 1,2} {\left( {c_{i\sigma ,\alpha }^\dag {c_{j\sigma ,\alpha }} + h.c.} \right)} } ,\\
H_\mathrm{int}& = U\sum\limits_i {{{\left( {\sum\limits_{\sigma ,\alpha } {{n_{i\sigma ,\alpha }}}-1 } \right)}^2}},
\end{align}
 where $\alpha =1,2$ denote two orbits. $c_{i\sigma ,\alpha}^{\dagger}$ ($c_{i\sigma ,\alpha}$ ) represents the electron creation (annihilation) in orbit $\alpha$ at the $i$th site with spin $\sigma$ ($\sigma=\uparrow,\downarrow$). $n_{i\sigma ,\alpha}$ is the electron number operator.
For each orbit, the honeycomb lattice on the torus or cylinder is spanned by length vectors $\mathbf{L_x}=L_x \mathbf{e_x}$ and $\mathbf{L_y}=L_y\mathbf{e_y}$, where $\mathbf{e_x}=(1,0)$ and $\mathbf{e_y}=(1/2, \sqrt{3}/2)$  are two primitive vectors, then the total number of sites for each orbit is $N_0=L_x\times L_y\times 2$ , where $L_x$ and  $L_y$ represent the number of unit cells along the two primitive-vector directions.
For the two-orbital system on honeycomb lattice we have the number of sites  $N=2\times N_0$ (where $2$ denotes two orbits)
and the number of electrons $N_e=N_0$  for a quarter filling.

We set $t$ as the unit of energy and consider Coulomb repulsion $U>0$. Given the additional sublattice and orbital degrees of freedom, we mainly focus on the cylinders or torus with circumferences up to $L_y=4$ unit cells ($2\times L_y=8$ lattice sites in each orbit).  In the present calculations, we keep up to $8000$ states with enough number of sweeps to get the converged data, the truncation error is of the order or less than $10^{-4}$. We also benchmark against quantum Monte Carlo (QMC)  for single orbital model\cite{SM}.

\begin{figure}[tbp!]
\begin{center}
\includegraphics[width=0.5\textwidth]{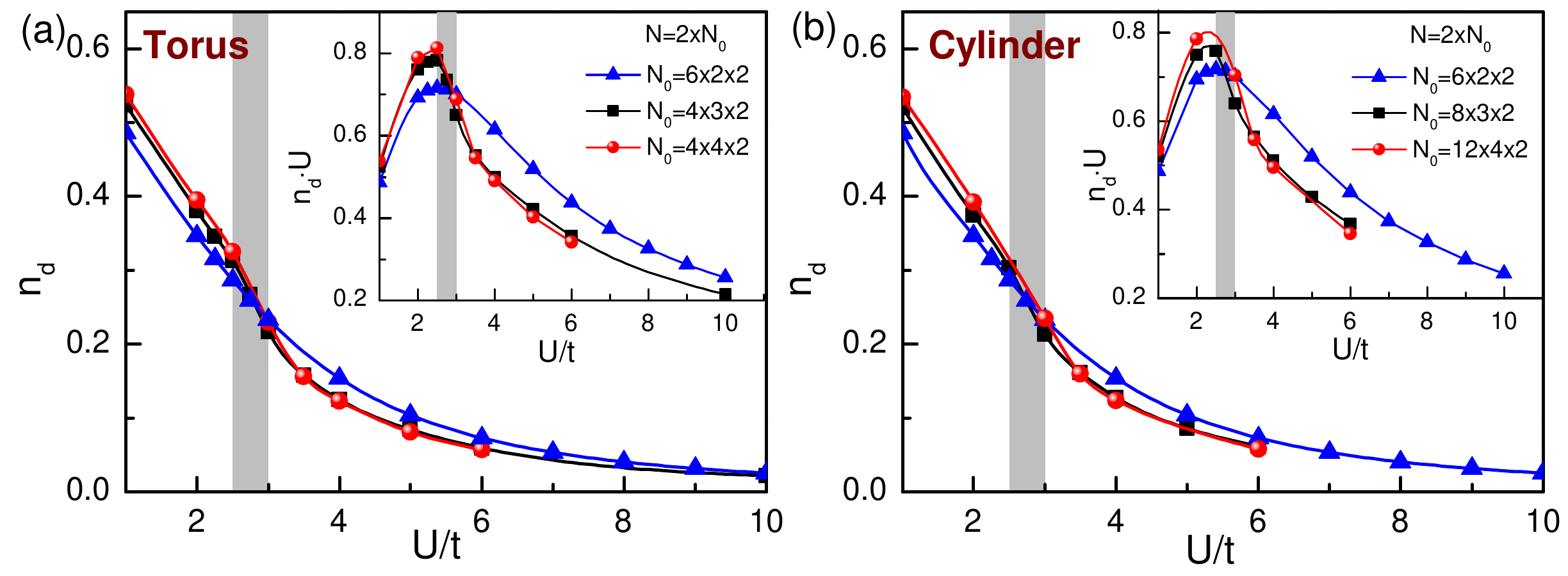}
\end{center}
\par
\renewcommand{\figurename}{Fig.}
\caption{ The average of the double occupancy $n_d$ as a function of $U/t$ for torus (a) and cylinder(b) geometries. The insets show the $n_d\cdot U$ as a function of $U/t$. The shadow bar indicates  the MIT. }
\label{Fig:D}
\end{figure}

\begin{figure}[tbp]
\begin{center}
\includegraphics[width=0.5\textwidth]{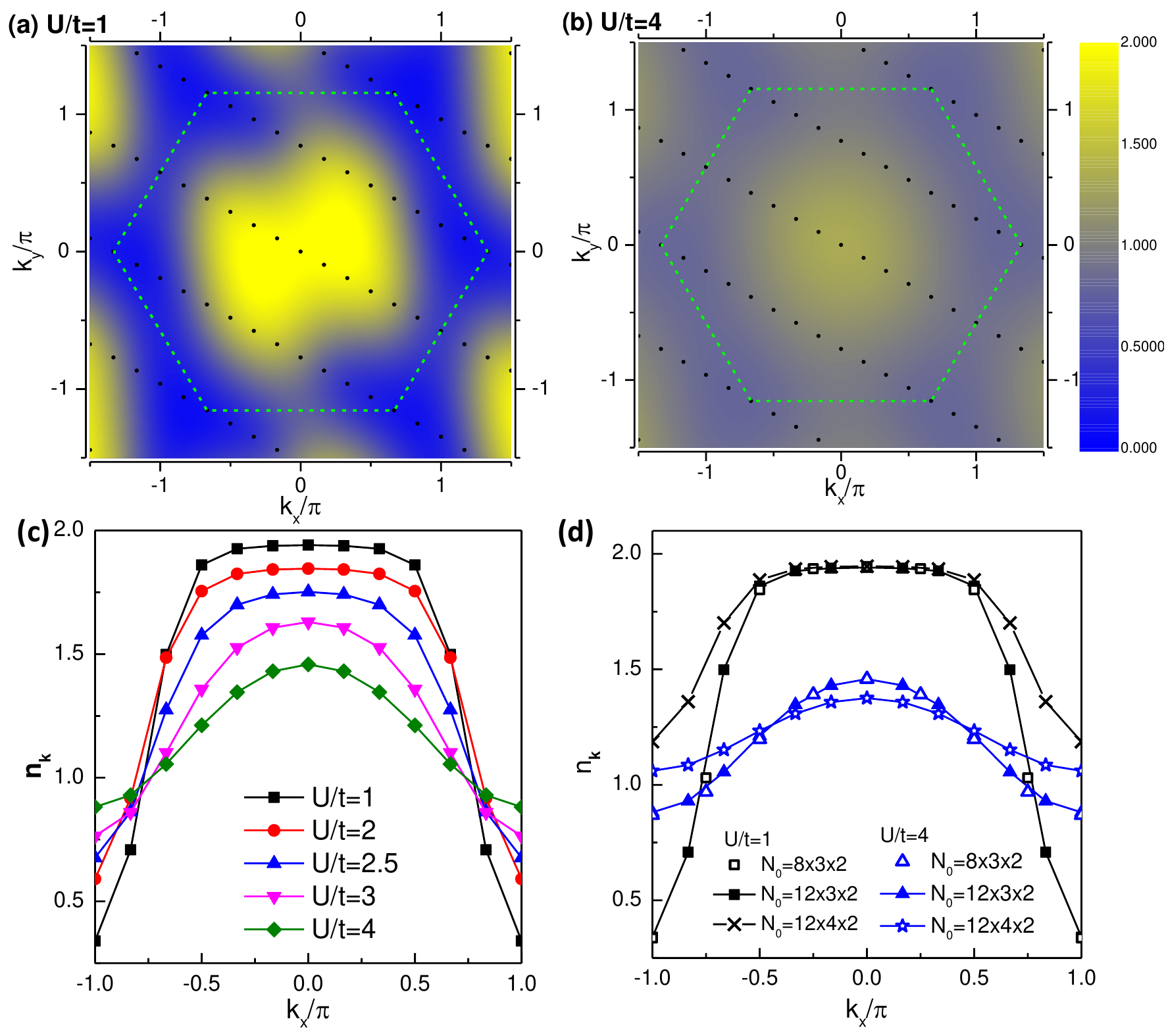}
\end{center}
\par
\renewcommand{\figurename}{Fig.}
\caption{The momentum distribution $n(\mathbf{k})$ for the metallic phase at $U/t=1$ (a) and insulating phase at $U/t=4$ (b).  The black dots represent the momentum points we can access in the  Brillouin zone of the finite size lattice, the contour plot is created by using triangulation interpolation. The cut of $n(\mathbf{k})$ along the line crossing $\Gamma$ point for different ratios of $U/t$ are shown in (c), where the slope of $n(\mathbf{k})$ decreases with the increase of $U/t$ near the Fermi surface. Here, we consider the cylinders with $N_0=12\times3\times2$. (d) The same cut of $n(\mathbf{k})$ as (c) at $U/t=1$ and $U/t=4$ for different system sizes.}
\label{Fig:nk}
\end{figure}
\begin{figure*}[tbp]
\begin{center}
\includegraphics[width=0.9\textwidth]{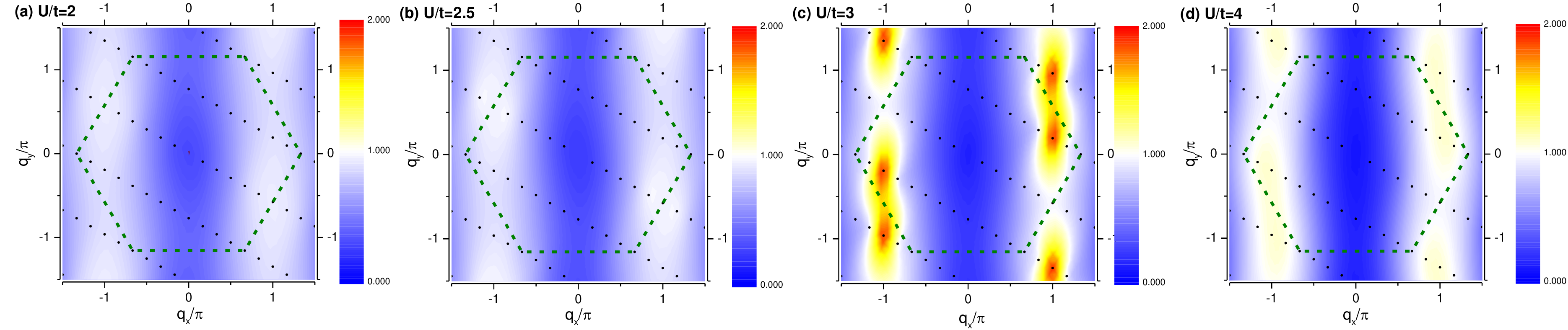}
\end{center}
\par
\renewcommand{\figurename}{Fig.}
\caption{The contour plot of the structure factor $N_\mathbf{q}$ for $L_y=3$ cylinders at $U/t=2$ (a) , and $U/t=2.5$ (b) ,  $U/t=3$ (c), $U/t=4$ (d).  $N_\mathbf{q}$ are featureless for both metallic phase (a) and insulating phase (d), while the CDW fluctuations arise in the vicinity of the MIT (c). The splitting of the CDW peaks around $M$ point signifies the incommensurate CDW instability. Here, we consider the cylinders with $L_x=12$ unit cells, the black dots represent the momentum points we can access in the  Brillouin zone of the finite size lattice. The contour plot is created by using triangulation interpolation. }
\label{Fig:Nq}
\end{figure*}
\begin{figure*}[tbp]
\begin{center}
\includegraphics[width=0.9\textwidth]{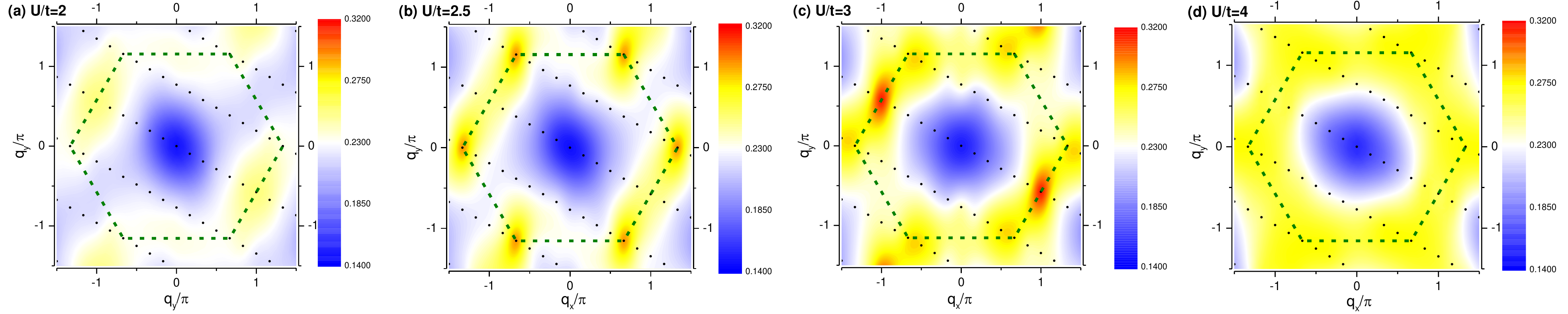}
\end{center}
\par
\renewcommand{\figurename}{Fig.}
\caption{The contour plot of the static spin structure factor $S_\mathbf{q}$ for $L_y=3$ cylinders at $U/t=2$ (a), $U/t=2.5$ (b),  $U/t=3$ (c), $U/t=4$ (d). $S_\mathbf{q}$ are featureless for both metallic phase (a) and insulating phase (d), while the spin/orbital density wave  fluctuations arise in the vicinity of the MIT point (b-c). The peaks of $S_\mathbf{q}$ locate at  $K$ or $M$ points indicating the commensurate spin/orbital density wave instability. Here, we consider the cylinders with $L_x=12$ unit cells, the black dots represent the momentum points we can access in the  Brillouin zone of the finite size lattice. The contour plot is created by using triangulation interpolation. The peaks in (c) is  higher than (b), indicating the stronger fluctuations with wave vector at $M$ points than at $K$ points. }
\label{Fig:Sq}
\end{figure*}

 \emph{Metal-Insulator Transition.}---We begin with studying the expectation value of the double occupancy $n_d$
 $ n_d = \frac{1}{{{N_0}}}\sum_{i = 1}^{{N_0}} {\left\langle {{{\left[ {\sum_{\sigma ,\alpha } {{n_{i\sigma ,\alpha }}}  - 1} \right]}^2}} \right\rangle }$,
which corresponds to the first order derivative of the ground-state energy and is a good measurement for the Mott transition. Figures~\ref{Fig:D} (a) and (b) show $n_d$ versus $U/t$ for the model Hamiltonian  (\ref{Model}) on torus and cylinder. With the increase of $U/t$,  $n_d$  monotonically decreases, which eventually  freezes the charge degrees of freedom.
 The kinks of each curve and  crossing of different curves with different lattice sizes indicate the transition takes place near $U_c/t=2.5\sim3$, which are independent of  lattice geometries we employed.
We also show $U\cdot n_d$  as a function of  $U/t$ in the insets of Fig.~\ref{Fig:D}, where the peaks in the curves may characterize the transition since  the derivatives of the curve show discontinuity from both sides of the  maximum point.

To  establish  the nature of  each phase,
 we examine the momentum distribution function $n(\mathbf{k})$, which is defined by the Fourier transformation of single particle propagator $\left\langle {c_{i\sigma }^\dag {c_{j\sigma }}} \right\rangle $, i.e., $n\left( \mathbf{k} \right) = \frac{1}{{{N_0}}}\sum\limits_{i,j,\sigma } {\left\langle {c_{i\sigma }^\dag {c_{j\sigma }}} \right\rangle {e^{i{\mathbf{k}}({{\bf{r}}_{\bf{i}}} - {{\bf{r}}_{\bf{j}}})}}}$ . As shown in  Figure~\ref{Fig:nk} (a) and (b),  $n(\mathbf{k})$ shows distinct behavior before and after the transition.
For the metallic phase at $U<U_c$, there is a sudden drop  near the Fermi surface [see Fig.~\ref{Fig:nk} (a)], while $n(\mathbf{k})$ is near flat without any singularity in the insulating side  at $U>U_c$ [see Fig.~\ref{Fig:nk} (b)]. To show the evolution of the Fermi surface with increasing $U/t$, we cut the $n(\mathbf{k})$ along the line crossing $\Gamma=(0,0)$ point, as shown in Fig.~\ref{Fig:nk} (c), the smooth change of the line shape indicates the continuity of this MIT, which is consistent with the behavior
of  $n_d$ [see Fig~\ref{Fig:D}]. In addition, We also confirm that results of $n(\mathbf{k})$ are robust against system size as illustrated in Fig.~\ref{Fig:nk} (d).

 \emph{Density Wave Fluctuations/Orders.}--- We have identified that the MIT occurs near  $U_c/t=2.5\sim3$. Below we examine the electronic and magnetic fluctuations and explore possible orders particularly near the MIT and in the insulating phase. Numerically the most direct evidence to explore the  fluctuations or orders is to calculate the correlations and their structure factors. Here, we consider the correlations within the same sublattice, which are the same for the two sublattices. Although the systems on cylinders break  translational symmetry due to the open boundaries,
we find our results are robust by neglecting the boundary effects  when the system size is relatively large.

\begin{figure}[tpb]
\begin{center}
\includegraphics[width=0.5\textwidth]{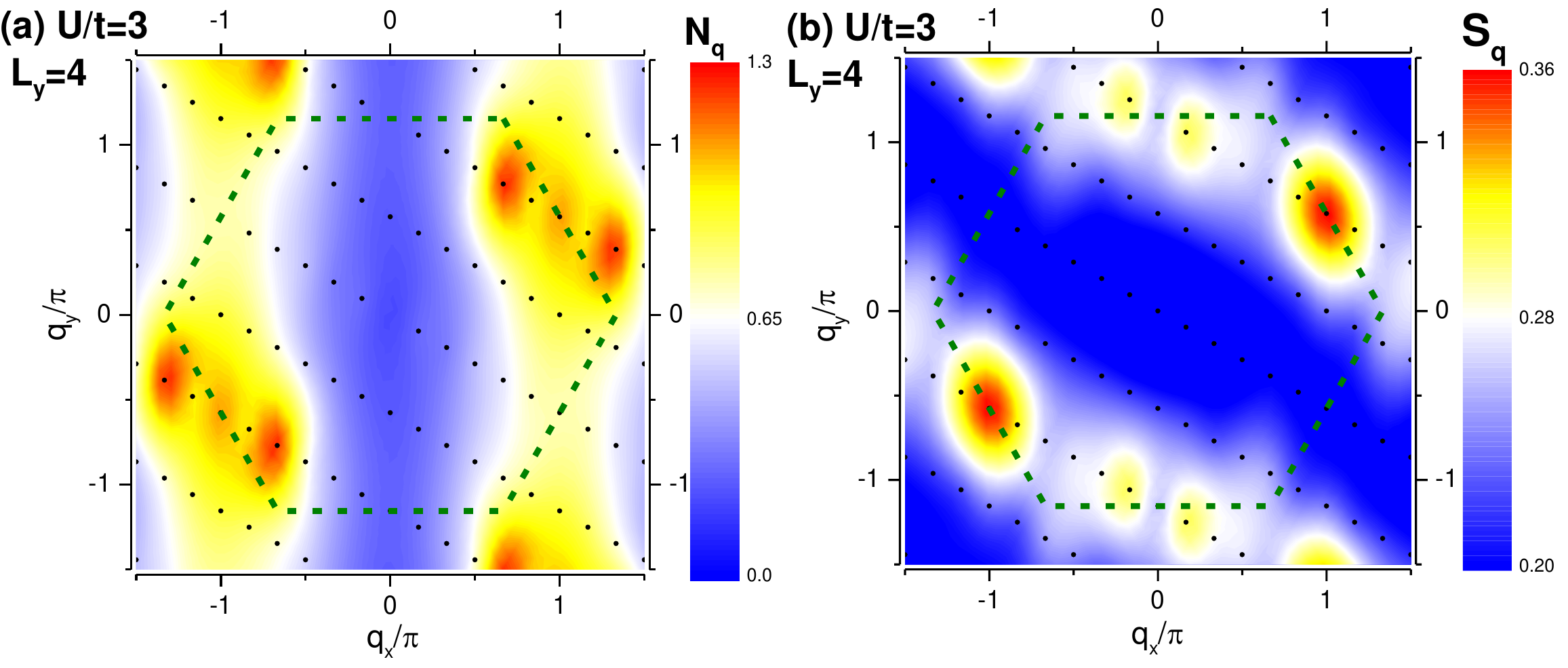}
\end{center}
\par
\renewcommand{\figurename}{Fig.}
\caption{The structure factor $N_\mathbf{q}$ (a) and $S_\mathbf{q}$ (b) for $L_y=4$ systems at $U/t=3$. Here, we consider the cylinders with $L_x=12$, the black dots represent the momentum points we can access in the  Brillouin zone of finite size lattice system. The contour plot is created by using triangulation interpolation.}
\label{Fig:Scaling}
\end{figure}

We first measure the structure factor of charge density-density correlations $ N_\mathbf{q} =\frac{1}{N_0}\sum\limits_{i,j} {\left\langle {n_i n_j} \right\rangle e^{i \mathbf{q}(\mathbf{r_i}-\mathbf{r_j})}}$.
Figures ~\ref{Fig:Nq} shows $N_q$ in the metallic phase, insulating phase as well as around the vicinity of the MIT. The black dots indicate the data points in the contour plot for finite size system with $L_x=12$ and $L_y=3$. Here we have substracted the peak at $q=0$ , which is trivially induced by the uniform charge background. Fig.~\ref{Fig:Nq} (a) and (d) show that there are no significant peaks in $N_\mathbf{q}$ deep inside both the metallic phase at $U<U_C$  and insulating phase at $U>U_C$, while near the transition point, $N_\mathbf{q}$ displays strong charge density wave (CDW) fluctuations, as shown in Fig.~\ref{Fig:Nq} (c). When $U/t=3$, the peaks in $N_q$ locate at  available momenta
nearby  $\mathbf{M}$ points, implying the incommensurate nature of such CDW fluctuations. We also check the finite size effect by studying  wider cylinder with $L_y=4$, as shown in the Fig.~\ref{Fig:Scaling} (a).  $N_q$ also displays significant peaks around $\mathbf{M}$ but with different wave vectors, which is consistent with the fact that the wave vectors of the incommensurate CDW  fluctuations depend  on the system geometry. Meanwhile, we also notice that  the intensity of $N_q$ is weakened with increasing the width of cylinders, as shown in  Fig.~\ref{Fig:Nq} (c) and Fig.~\ref{Fig:Scaling} (a), which suggests the CDW fluctuations are not strong enough to form CDW order  towards 2D limit.

In the spin channel,  we study the static spin structure factor to detect the magnetic fluctuations, which is defined as the Fourier transformation of spin-spin correlations, i.e., $S_\mathbf{q} =\frac{1}{N_0}\sum\limits_{i,j} {\left\langle {S_i^zS_j^z} \right\rangle e^{i \mathbf{q}(\mathbf{r_i}-\mathbf{r_j})}}$,
where $S^a_i \equiv \sum_{\alpha} \sum_{\sigma \sigma^\prime} c^\dagger_{i\sigma, \alpha} s^a_{\sigma \sigma^\prime} c_{i\sigma^\prime,\alpha}$ is the
spin operator on site $i$.
Due to the $SU(4)$ symmetry of the Hubbard model, the orbital and spin correlations are identical.
Figures ~\ref{Fig:Sq} show the spin/orbital structure factor $S_\mathbf{q}$ for different values of $U/t$ for the same $L_y=3$ systems with $L_x=12$.  Both the metallic phase at $U<U_C$ [see Fig.~\ref{Fig:Sq} (a)] and the insulating phase at $U>U_C$ [see Fig.~\ref{Fig:Sq} (d)] exhibit  the absence of magnetic fluctuations or orders. While near $U_c$,  $S_\mathbf{q}$ displays strong SODW fluctuations, as shown in  Fig.~\ref{Fig:Sq} (b) and  Fig.~\ref{Fig:Sq} (c). Interestingly, the SODW  fluctuations display competing wave vectors at commensurate momentum  $\mathbf{K}$ or $\mathbf{M}$, which depend on the values of ratio between Coulomb interaction and bandwidth. When $U/t=2.5$ [see Fig.~\ref{Fig:Sq}(c)], there are peaks located at $\mathbf{K^+}=(4\pi/3,0)$ and $\mathbf{K^-}=(2\pi/3,2\pi/\sqrt 3)$ points in the Brillouin zone.
However, further increasing the ratio to $U/t=3$ [see Fig.~\ref{Fig:Sq}(d)], the peaks of $S(q)$ near $\mathbf{K^\pm}$ become broader while much sharper peaks arise at $M=(\pm \pi,\mp\pi/\sqrt 3)$ points,  indicating the  SODW  fluctuations with wave vector $\mathbf{M}$ are more dominant than  the ones with wave vector $\mathbf{K}$ near $U_c$
This can be seen more clearly by increasing the  cylinder width towards 2D, as shown in the Fig.~\ref{Fig:Scaling} (b) for $S_q$ on $L_y=4$ cylinders,  where the SODW  fluctuations display significant peaks around $\mathbf{M}$. In addition, the intensity of $S_q$ is also enhanced with increasing the system size [see Fig.~\ref{Fig:Sq} (c) and Fig.~\ref{Fig:Scaling} (b)], indicating the possible existence of SODW order in 2D.

\begin{figure}[tpb]
\begin{center}
\includegraphics[width=0.35\textwidth]{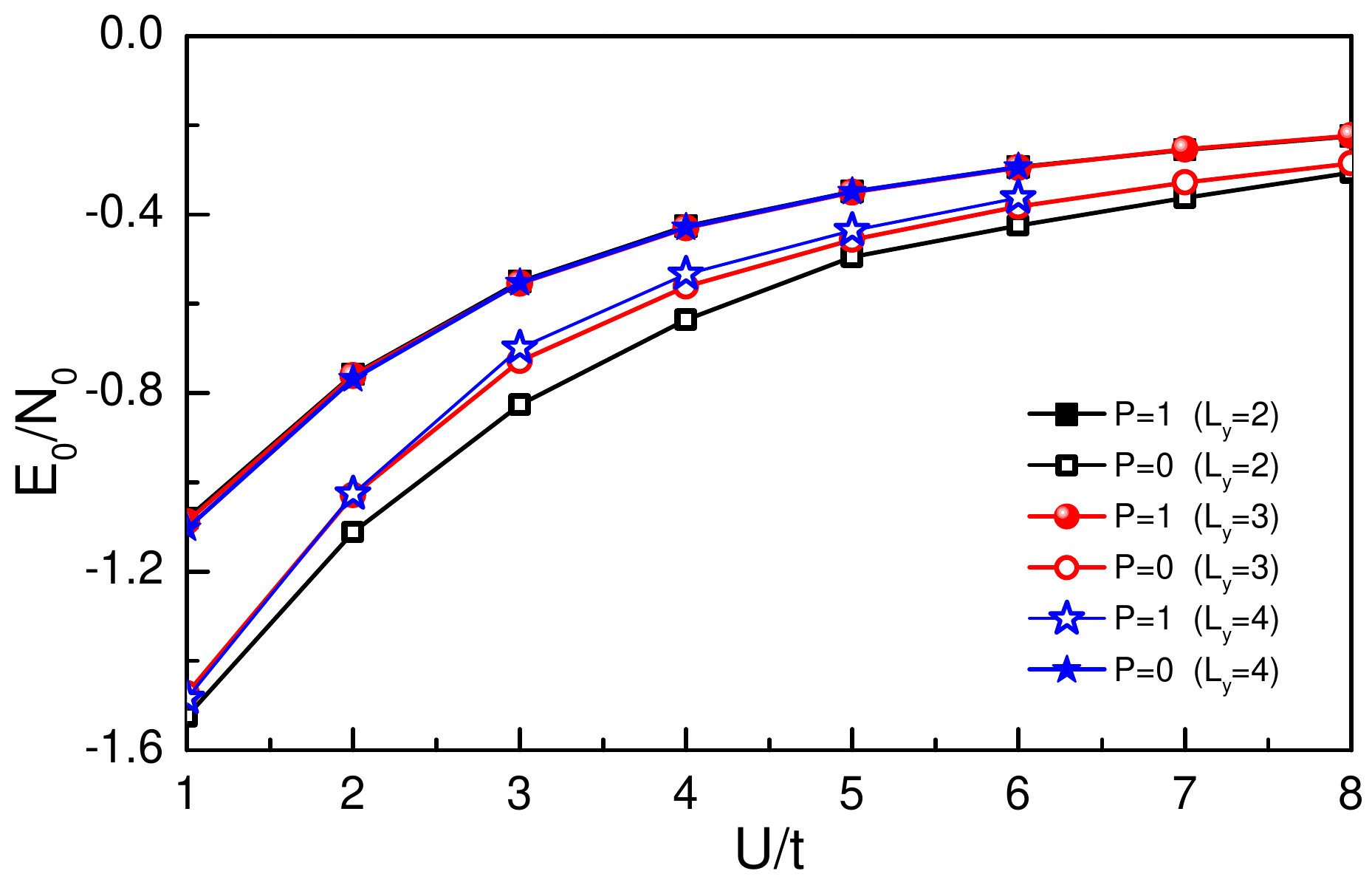}
\end{center}
\par
\renewcommand{\figurename}{Fig.}
\caption{The ground state energy $E_0$ per site as a function of $U/t$ for both unpolarized $P=0$ and polarized $P=1$ sectors. Here, we consider the systems with $L_x=12$ and different $L_y$.}
\label{Fig:P}
\end{figure}

 \emph{Orbital Polarizations.}---We have identified strong charge and spin/orbital dentity wave  fluctuations or orders near the quantum phase transition, while
the structure factors indicate the featureless nonmagnetic insulating state at $U>U_c$ [see  Fig.~\ref{Fig:Nq}(d) and  Fig.~\ref{Fig:Sq}(d) ]. Below, we examine polarization in the spin/orbital channel.  To check the possibility of orbital polarizations of  the Hamiltonian  [see Eq.~(\ref{Model})] at a quarter filling, we define $P=|N_1-N_2|/N_0$, where $N_\alpha$ ($\alpha=1,2$) represent the total electron number in orbit $\alpha$, and then we compare the ground-state energies of the unpolarized sector with equal number of electrons in two orbits ($P=0$) and the orbital polarized sector with all electrons in one orbit ($P=1$).  As shown in the Fig.~\ref{Fig:P} for different $L_y$ systems, $P=0$ sector always has lower energy than $P=1$ sector, which is independent of  the system sizes, implying the absence of the fully orbital polarization for the ground states. Due to the $SU(4)$ symmetry, this result also implies that the ground state is not fully spin-polarized. This conclusion is consistent with the absence of enhancement in the $\mathbf{q}=0$ spin structure factor.

 \emph{Discussion and Summary.}---We now discuss the relevance of our work to real materials. Recently, the two-orbital $SU(4)$ Hubbard model on the honeycomb lattice at quarter filling has been proposed as a realistic model for $\alpha$-ZrCl$_3$ and metal organic frameworks with tricoordinated lattices \cite{Yamada}. It may be possible to tune the ratio $U/t$ by pressure and to search for spin/orbital density wave and Mott insulator states.

For TBG, the microscopic two-orbital Hamiltonian derived from band structure calculations and Coulomb interactions projected to low-energy bands has more terms than the simplest two-orbital Hubbard model studied here. It includes single-particle hoppings beyond nearest neighbors\cite{Senthil}, extended density and exchange interactions beyond on-site $U$,
pair hoppings \cite{Vafek,Guinea}, and the effect of other Moir\'{e} bands may be important~\cite{Balents, Bernevig, Vishwanath}.
 These additional terms can stabilize SODW\cite{liang-newer}  or lead to new phases beyond the density wave and Mott insulator states. For example, Ref.\cite{Vafek2} found a orbital-polarized state at quarter filling induced by pair hopping, when kinetic energy is set to zero.
The fully understanding the physics in TBG requires a more comprehensive study of the microscopic Hamiltonian. Nonetheless, the current two-orbital Hubbard model with the nearest-neighbor hopping and on-site repulsion may be served as a starting and reference  point to study correlated electron physics of TBG.

In summary, based on DMRG simulations of the two-orbital Hubbard model on honeycomb lattice, we identify a MIT near $U_c/t\approx 2.5\sim3$  accompanied by commensurate SODW orders and incommensurate CDW fluctuations, and find a nonmagnetic Mott insulator without orbital polarization at $U>U_c$.
Our results enrich the Mott physics near MIT, where a SODW  emerges as an intermediate phase. We also find the SODW  is robust against lightly doping and the weak nearest neighbor (NN) interactions\cite{SM,Wehling2011}. In addition,
we have also checked that there is no indication of any density waves at 1/8 filling\cite{SM} , which corresponds to one electron per unit cell. Our findings can be tested experimentally in TBG, where applying pressure can drive the system across MIT and deep into the Mott insulator \cite{Yankowitz2018}.

\begin{acknowledgments}
We thank A. Vishwanath, S. K. Jian and Masahiko Yamada for helpful discussions.
This research was carried out through the support of the David and Lucile Packard foundation (Z.Z., L.F.).
This work  carried out at CSUN was  supported by the U.S. Department of Energy (DOE), Office of Basic Energy Sciences under Grant No. DEFG02-06ER46305 (Z.Z., D.N.S.).
D.N.S. also acknowledges  support from the DOE, the Office
of Basic Energy Sciences, Division of Materials Sciences
and Engineering, under Contract No. DE-AC02-76SF00515
through SLAC National Accelerator Laboratory
for the development of numerical methods extensively applied
to this project.
\end{acknowledgments}

\end{document}